\documentclass{ws-p8-50x6-00}

\begin{document}

\title{Recent Results from the HDMS Experiment}

\author{H.V. Klapdor-Kleingrothaus, L. Baudis, A.~Dietz, G.~Heusser, 
I.V. Krivosheina, B.~Majorovits, St.~Kolb, H.~Strecker}

\address{Max Planck Institut f\"ur Kernphysik, P.O. Box 103980, 69029 
Heidelberg, Germany, Home Page Heidelberg Non-Accelerator 
Particle Physics group: http://mpi-hd.mpg.de.non$\_$acc/}

\maketitle

\abstracts{
The status of dark matter search with the HDMS experiment is reviewed. 
After one
year of running the HDMS prototype detector in the Gran Sasso 
Underground Laboratory, the inner crystal of the detector has been
replaced with a HPGe crystal of enriched $^{73}$Ge.  The results of
the operation of the HDMS prototype detector are discussed.
}

The Heidelberg Dark Matter Search (HDMS) Experiment is designed to look
for Weakly Interacting Dark Matter Particles (WIMPs). 
Through a special configuration
an effective background reduction is achieved with respect to a 
conventional design, which results in an increase
of sensitivity for WIMP Dark Matter.

Here we present first results of the Heidelberg
Dark Matter Search (HDMS) experiment \cite{prophdms,hdms}, whose
prototype took data over a period
of about 15 months in the Gran Sasso Underground Laboratory at LNGS in Italy.
The last 49 days of data taking
are analyzed in terms of WIMP-nucleon cross sections and a
comparison to other running dark matter experiments is made. 
The final experiment operating an enriched $^{73}$Ge crystal inside a
natural Germanium detector has been installed in august 2000.

\section{The HDMS experiment}
\begin{figure}
\epsfysize=10pc
\epsfbox{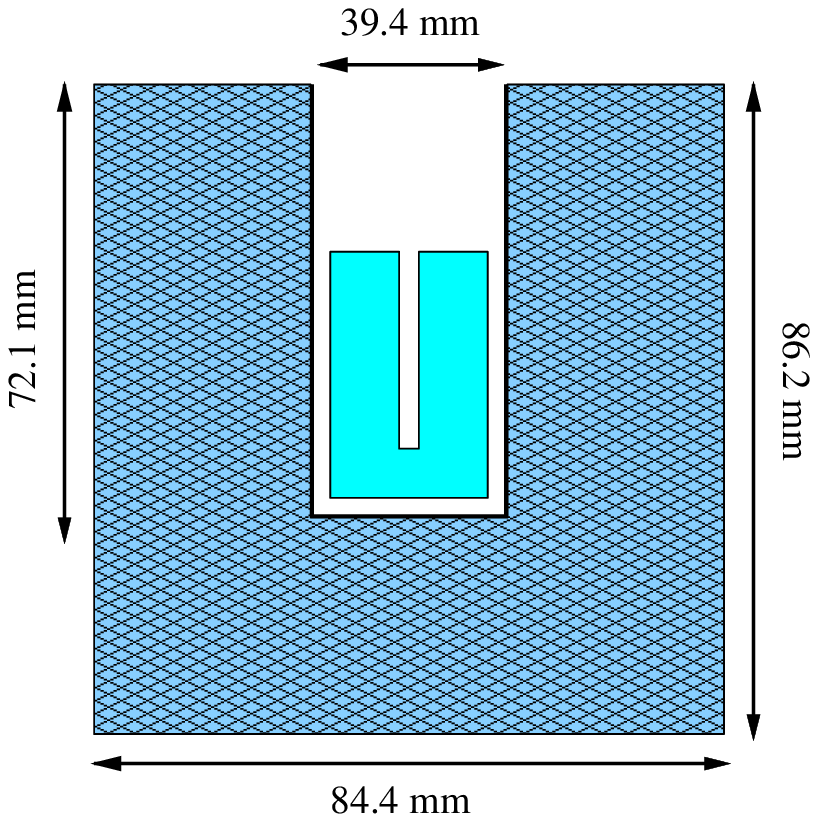}
\epsfysize=10pc
\hspace*{0.5cm}
\epsfbox{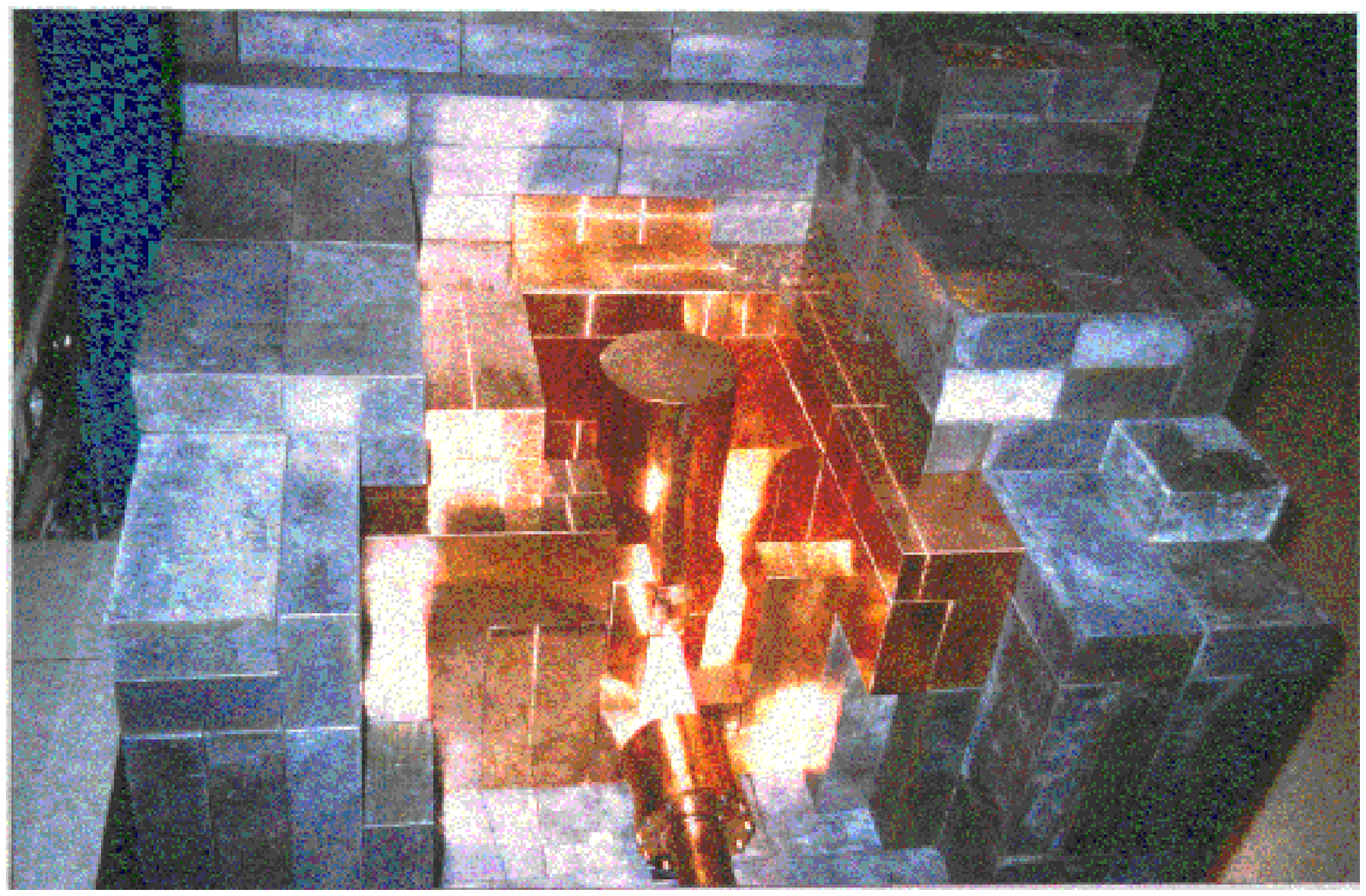}
\caption[]{Left: Schematic view of the HDMS experiment. A
  small Ge crystal is surrounded by a well type Ge-crystal,
  the anti-coincidence between them is used to suppress background
  created by external photons. Right: The HDMS prototype detector
  during its installation at LNGS.} 
\label{detindet}
\end{figure}

HDMS operates two ionization HPGe detectors in a unique configuration
\cite{prophdms,hdms}.
A small, p-type Ge crystal is surrounded by a well-type Ge crystal,
both being mounted into a common cryostat system (see
Figure~\ref{detindet} for a schematic view). 
Two effects are expected to reduce the background of the
inner target detector with respect to our best measurements with the
Heidelberg-Moscow experiment \cite{WIMPS}. First, the anticoincidence
between the two detectors acts as an effective suppression of multiple
scattered photons. Second, we know that the main radioactive
background of Ge detectors comes from materials situated in the immediate
vicinity of the crystals. In the case of HDMS the inner detector is
surrounded (apart from the thin isolation) by a second Ge crystal -
one of the radio-purest known materials.

In order to house both Ge crystals and to establish the two
high-voltage and two signal contacts, a special design of the copper
crystal holder system was required. 
The cryostat system was built in Heidelberg and made of low
radioactivity copper, all surfaces being electro-polished.
The FETs are placed 20 cm away from the crystals, their effect on the
background is minimized by a small solid angle for viewing the
crystals and by 10 cm of copper shield.

\subsection{Detector Performance at LNGS}

The HDMS prototype was installed at LNGS in March 1998.
Figure \ref{detindet} shows the detector in its open shield. The
inner shield is made of 10 cm of electrolytic copper, the outer
one of 20 cm of Boliden lead.  The whole setup is enclosed in an air
tight steel box and flushed with gaseous nitrogen in order to suppress
radon diffusion from the environment. Finally a 15 cm thick borated
polyethylene shield surrounds the steel box in order to minimize the
influence of neutrons from the natural radioactivity and muon produced
neutrons in the Gran Sasso rock.

The prototype detector successfully
took data over a period of about 15 months, until July 1999.
The individual runs were about 0.9 d long. Each day the experiment
was checked and parameters like leakage current of the detectors,
nitrogen flux, overall trigger rate and count rate of each detector
were checked. The experiment was calibrated weekly with a $^{133}$Ba
and a $^{152}$Eu-$^{228}$Th source. 
The energy resolution of both detectors (1.2\,keV at 300\,keV inner
detector and 3.2\,keV at 300\,keV outer detector) were stable as a function of time.
The zero energy resolution is 0.94\,keV for the inner detector and 3.3
keV for the outer one.

The energy thresholds are 2.0\,keV and 7.5\,keV for the inner and outer
detector, respectively.

\begin{figure}[t]
\hspace*{-0.6cm}
\epsfxsize=15pc
\epsfbox{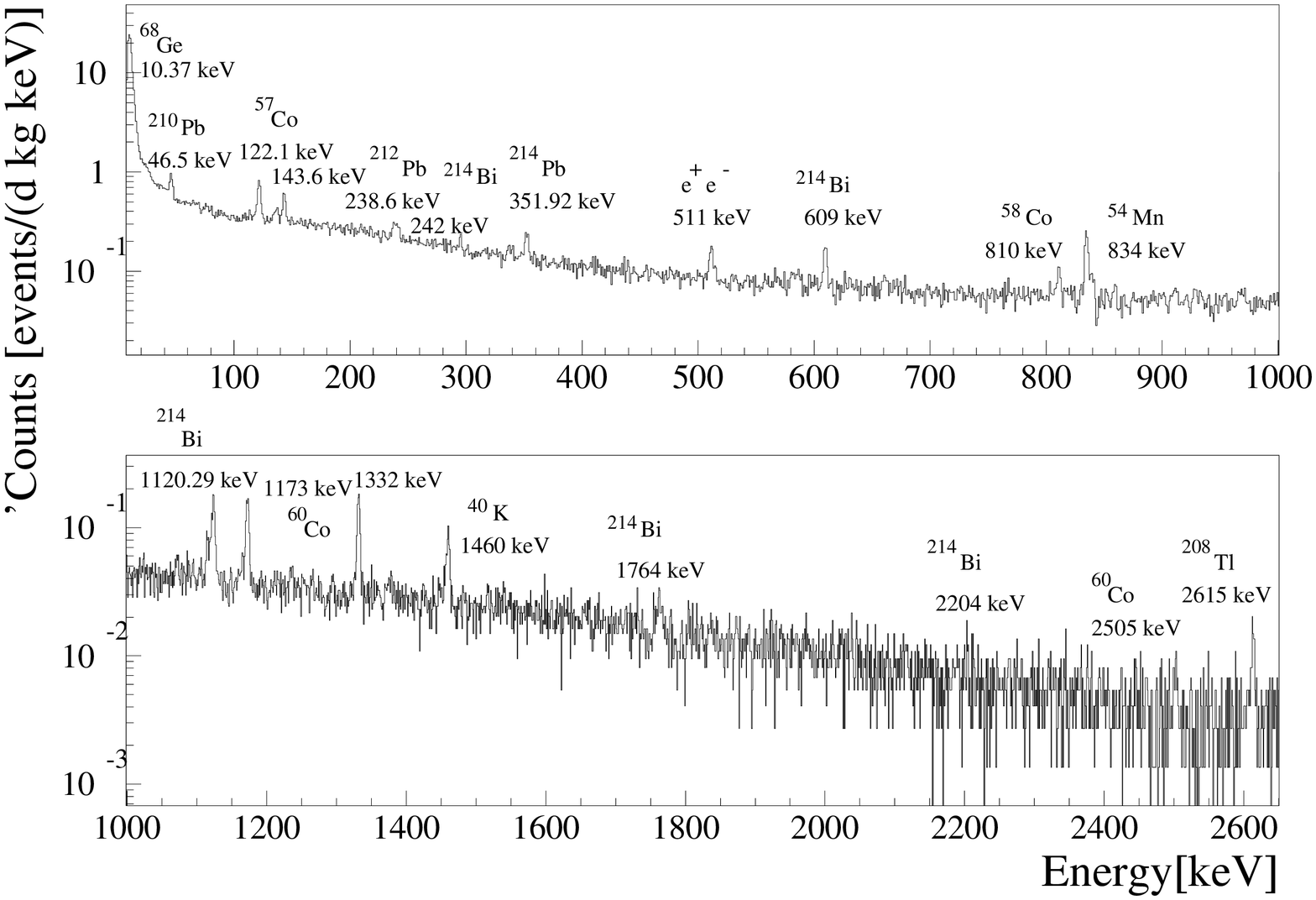}
\epsfxsize=15pc
\epsfbox{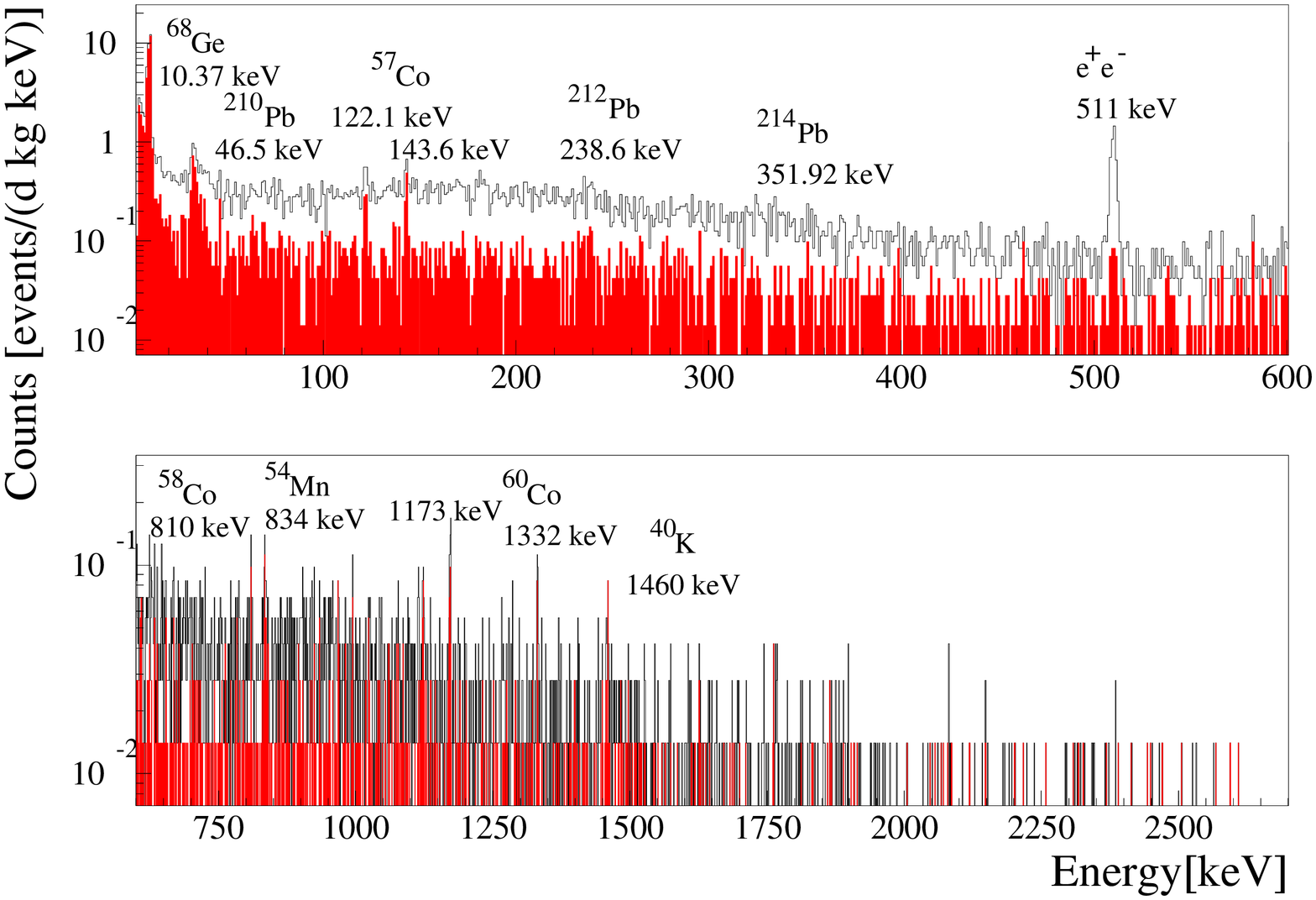}
\caption[]{Left: Sum spectrum of the outer detector after a
  measuring time of 363 days. The most prominent lines are labeled.
  Right: Sum spectrum of the inner detector after a
  measuring time of 363 days. The most prominent lines are
  labeled. The filled histogram is the spectrum after the
  anti-coincidence with the outer detector.
}
\label{sum-outer} 
\end{figure}

Due to the very special detector design, we see a cross-talk
between the two detectors. The observed correlation is linear and can
be corrected for off-line \cite{dm98}.
After correction for the cross talk and recalibration to standard
calibration values, the spectra of the daily runs were summed. Figure
\ref{sum-outer} shows the sum spectra for the
outer and inner detector, respectively (the most important identified 
lines are labeled).

In the outer detector lines of some cosmogenic and anthropogenic
isotopes, the U/Th  natural decay chains and $^{40}$K are clearly
identified.
The statistics in the inner detector is not as
good, however the X-ray at 10.37\,keV resulting from the decay of
$^{68}$Ge, some other cosmogenic isotopes, $^{210}$Pb and  $^{40}$K
can be seen.
The region below 10\,keV is dominated by the X-rays from cosmogenic
radio nuclides. In addition a
structure centered at 32\,keV is identified. Its
origin is not yet clear and is currently under investigation. 

After 363 days of pure measuring time the statistics in the inner
detector was high enough in order to estimate the background
reduction through the anti-coincidence with the outer detector.

\begin{figure}[t]
\epsfysize=10pc
\epsfbox{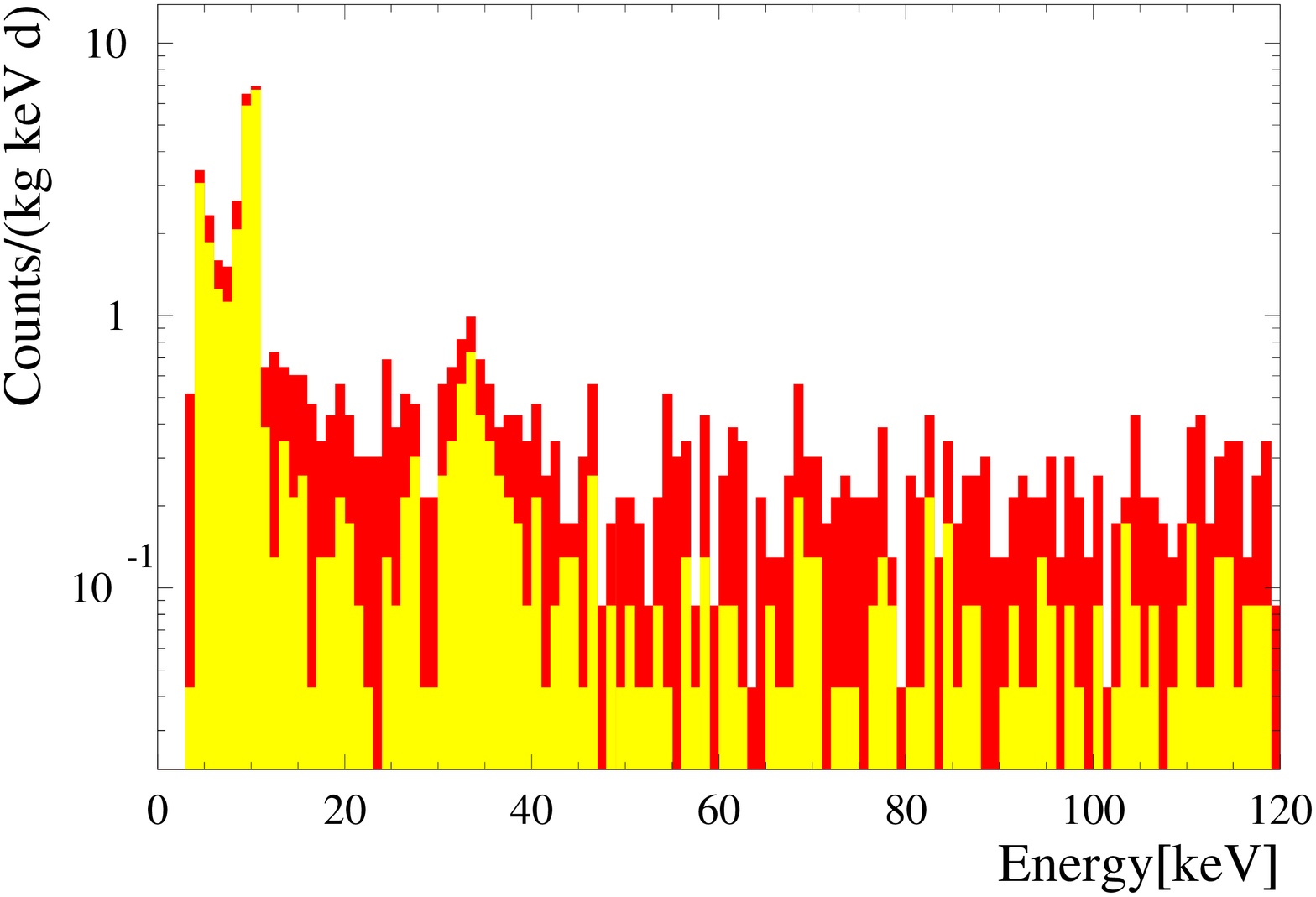}
\epsfysize=10pc
\epsfbox{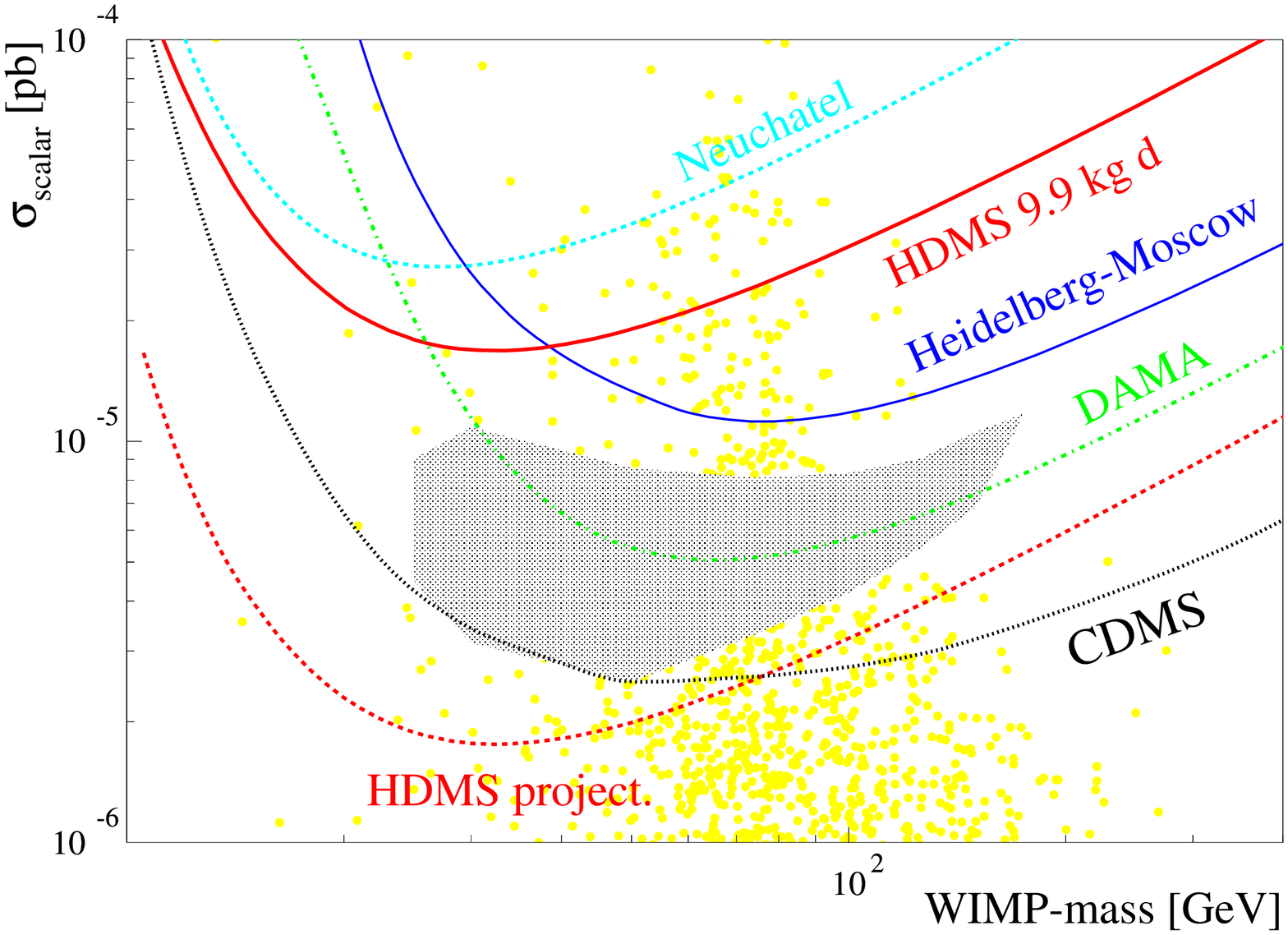}
\caption[]{Left: Low energy spectrum of the inner detector.
  The light shaded spectrum
  corresponds to the events using the anticoincidence, the dark shaded
  spectrum to all events.
Right:   WIMP-nucleon cross section limits as a function of the WIMP
  mass for spin-independent interactions. 
  The solid line corresponds to the limit set by the HDMS-prototype
  detector. 
  The other plain curves correspond to the limits given by CDMS
  \cite{rick2000}, 
  DAMA \cite{damaexcl} and the Heidelberg-Moscow Experiment \cite{WIMPS}.
  The
  filled contour represents  the 2$\sigma$ evidence region of the DAMA
  experiment \cite{damahere}. The dashed line corresponds to the
  expectation for the final HDMS detector, assuming a threshold of 2~keV.
}
\label{low-inner} 
\end{figure}

Figure \ref{low-inner} shows the low-energy spectrum of the inner detector
before and after the anti-coincidence. 
The cosmogenic X-rays below 11\,keV are preserved, since in this case
the decays are occuring in the inner detector itself. The same is
valid for the structure at 32\,keV, also a $^3$H spectrum with endpoint
at 18.6\,keV is presumably present.
If the anti-coincidence is evaluated in the energy region between
40\,keV and 100\,keV, the background reduction factor is 4.3. 
The counting rate after the anti-coincidence in this energy region is 
0.07\,events/(kg\,d\,keV), thus very close to the value measured in the 
Heidelberg-Moscow experiment with the enriched detector ANG2 \cite{WIMPS}.    
In the energy region between 11\,keV and 40\,keV the background index is 
with 0.2\,events/(kg\,d\,keV) a factor of 3 higher.

\subsection{Dark Matter Limits}

The evaluation for dark matter limits on the WIMP-nucleon cross section 
$\sigma_{\rm scalar}^{\rm W-N}$ follows the conservative 
assumption that the whole experimental spectrum consists of 
WIMP events.
Consequently, excess events from calculated 
WIMP spectra above the experimental spectrum in any energy region 
with a minimum width of the energy resolution of the detector 
are forbidden (to a given confidence limit). 

The parameters used in the calculation of expected WIMP spectra are 
summarized in \cite{WIMPS}. We use formulas given 
in the extensive review \cite{lewin} for a truncated 
Maxwell velocity distribution in an isothermal WIMP--halo 
model (truncation at the escape velocity, compare also \cite{freese}). 

After calculating the WIMP spectrum for a given WIMP mass, the scalar 
cross section is the only free parameter which is then used to fit the 
expected to the measured spectrum 
using a one-parameter maximum-likelihood fit algorithm. 

To compute the limit for the HDMS inner detector we took only the last 
49 days of measurement. We omit the first 260\,days in order to
reduce the contaminations due to long-lived cosmogenically produced
materials. These have life times of typically $\sim$~200 days.
The energy threshold of the measurement was 2\,keV.
%We took an energy threshold of 3\,keV (compared to the energy threshold 
%of 2.5\,keV for the inner detector).
The resulting preliminary upper limit exclusion plot in the 
$\sigma_{\rm{scalar}}^{\rm{W-N}}$ versus M$_{\rm{WIMP}}$
plane is shown in Fig.~\ref{low-inner}.

Already at this stage, the limit is competitive with our limit from
the Heidel\-berg-Moscow experiment. In the low mass regime for WIMPs the 
limit has been improved due to the low energy threshold of 2~keV
reached in this setup. 

Also shown in the figure are limits from the Heidelberg-Moscow
experiment \cite{WIMPS},
limits from the DAMA experiment \cite{damaexcl} and the most recent
results in form of an exclusion curve from the  
CDMS experiment \cite{rick2000}. The filled contour represents  the
2$\sigma$ evidence region of the DAMA experiment \cite{damahere}.

\subsection{Outlook for the HDMS Experiment and conclusions}

The prototype detector of the HDMS experiment successfully took data 
at LNGS over a period of about 15 months. 
Most of the background sources (with exception of the 32\,keV structure 
in the inner detector) were identified. The background reduction
factor in the inner detector through anticoincidence is about 4.
The background in the low-energy region of the inner
detector (with exception of the region still dominated by cosmogenic
activities) is already comparable to the most sensitive dark matter
search experiments.

For the final experimental setup, important changes were made.
The crystal holder was replaced by a holder made of ultra low
level copper, the soldering of the contacts was avoided, 
thus no soldering tin was used in the new setup and finally
the inner crystal made of natural Germanium in the described
prototype was replaced by an enriched $^{73}$Ge crystal. In this
way, the $^{70}$Ge isotope (which is the mother isotope for
$^{68}$Ge production) is
strongly de-enriched (the abundance in natural Germanium is 7.8\%).

After a period of test measurements in the low-level laboratory in
Heidelberg, the full scale experiment was installed at LNGS 
in August 2000. The energy threshold of the inner detector is unchanged
at 2.0~keV and the
energy resolution has slightly improved with respect to the prototype
detector \cite{diss}.

The projected final sensitivity of the detector can be 
seen in Fig.~\ref{low-inner}.

\end{document}